\documentclass[prb,aps,showpacs,twocolumn,superscriptaddress,amssymb,nobibnotes,epsf]{revtex4}

\usepackage{graphicx}
\usepackage{dcolumn}
\usepackage{bm}
\usepackage{tabularx}
\usepackage{amsmath}
\usepackage{textcomp}
\usepackage{upgreek}
\usepackage{SIunits}

\begin{document}

\title{Anomalous impurity effects in the iron-based superconductor KFe$_2$As$_2$}

\author{A. F. Wang}
\affiliation{Hefei National Laboratory for Physical Science at
Microscale and Department of Physics, University of Science and
Technology of China, Hefei, Anhui 230026, People's Republic of
China}

\author{S. Y. Zhou}
\affiliation{Department of Physics, State Key Laboratory of Surface
Physics, and Laboratory of Advanced Materials, Fudan University,
Shanghai 200433, China}

\author{X. G. Luo}
\affiliation{Hefei National Laboratory for Physical Science at
Microscale and Department of Physics, University of Science and
Technology of China, Hefei, Anhui 230026, People's Republic of
China}

\author{X. C. Hong}
\affiliation{Department of Physics, State Key Laboratory of Surface
Physics, and Laboratory of Advanced Materials, Fudan University,
Shanghai 200433, China}

\author{Y. J. Yan,$^{1}$ J. J. Ying,$^{1}$ P. Cheng,$^{1}$ G. J. Ye,$^{1}$ and Z. J. Xiang}
\affiliation{Hefei National Laboratory for Physical Science at
Microscale and Department of Physics, University of Science and
Technology of China, Hefei, Anhui 230026, People's Republic of
China}

\author{S. Y. Li}
\altaffiliation{Corresponding author} \email{shiyan_li@fudan.edu.cn}
\affiliation{Department of Physics, State Key Laboratory of Surface
Physics, and Laboratory of Advanced Materials, Fudan University,
Shanghai 200433, China}

\author{X. H. Chen}
\altaffiliation{Corresponding author} \email{chenxh@ustc.edu.cn}
\affiliation{Hefei National Laboratory for Physical Science at
Microscale and Department of Physics, University of Science and
Technology of China, Hefei, Anhui 230026, People's Republic of
China}

\begin{abstract}
High-quality K(Fe$_{1-x}$Co$_x$)$_2$As$_2$ single crystals have been
grown by using KAs flux method. Instead of increasing the
superconducting transition temperature $T_{\rm c}$ through electron
doping, we find that Co impurities rapidly suppress $T_{\rm c}$ down
to zero at only $x \approx$ 0.04. Such an effective suppression of
$T_{\rm c}$ by impurities is quite different from that observed
in Ba$_{0.5}$K$_{0.5}$Fe$_2$As$_2$ with multiple nodeless
superconducting gaps. Thermal conductivity measurements in zero field
show that the residual linear term $\kappa_0/T$ only change slightly with 3.4\%
Co doping, despite the sharp increase of scattering rate. The implications of
these anomalous impurity effects are discussed.

\end{abstract}

\pacs{74.70.Xa,74.25.fc,74.62.Dh}


\vskip 300 pt

\maketitle

The iron-based superconductors, as a second high-$T_{\rm c}$
superconducting family after cuprates, have stimulated great
interests.\cite{Hosono,Chen,JPaglione,GRStewart} Since the symmetry
and structure of the superconducting gap reflect the underlying
electron pairing mechanism, extensive works have been done to
clarify them.\cite{Hirschfeld} However, so far there is still no
consensus on this issue, particularly the origin of the nodal gaps in
KFe$_2$As$_2$, \cite{Fukazawa,LiSY,Hashimoto1,Reid,KOkazaki,Tafti} LaFePO, \cite{Fletcher,Hicks,MYamashida}
LiFeP, \cite{KHashimoto2} BaFe$_2$(As$_{1-x}$P$_x$)$_2$, \cite{YNakai,KHashimoto3,ZhangY}
and Ba(Fe$_{1-x}$Ru$_x$)$_2$As$_2$. \cite{XQiu}

For the hole-doped Ba$_{1-x}$K$_x$Fe$_2$As$_2$ (``122") system, the
angle-resolved photon emission spectroscopy (ARPES) expeirment
revealed multiple nodeless gaps on both hole and
electron Fermi surfaces at optimal doping $x = 0.4$,\cite{HDing} which was further supported by measurements of bulk
properties such as thermal conductivity.\cite{LuoXG} On the
overdoped side, nodeless gaps were also observed at $x$ = 0.7 by
ARPES,\cite{HDing2} and at $x$ = 0.77 by point contact Andreev
reflection spectroscopy.\cite{ZhangXH} Surprisingly, further
increasing doping to the end member KFe$_2$As$_2$, in which only hole pockets exist,\cite{KFeAs} nodal
superconducting gap was found by NMR, thermal conductivity, penetration
depth, and ARPES measurements.\cite{Fukazawa,LiSY,Reid,Hashimoto1,KOkazaki}

While the $s_{\pm}$-wave gaps in Ba$_{0.6}$K$_{0.4}$Fe$_2$As$_2$ are
still not conclusive,\cite{Hirschfeld} it is under hot debate
whether the superconducting gap of KFe$_2$As$_2$ is $d$-wave with
symmetry-imposed nodes \cite{Reid,Thomale} or $s$-wave with accidental
nodes.\cite{KOkazaki,Suzuki,Maiti} The thermal conductivity experiment
showed evidences for $d$-wave superconductivity by comparing the results of
dirty and clean KFe$_2$As$_2$ single crystals,\cite{Reid} however, the ARPES experiment clearly
observed eight nodes on the middle Fermi surfaces, suggesting accidental
nodal $s$-wave superconducting gap in KFe$_2$As$_2$.\cite{KOkazaki}

Recently, the heavily hole-doped Ba$_{1-x}$K$_x$Fe$_2$As$_2$ single crystals are available,
and more ARPES and thermal conductivity experiments have been done to figure out the doping evolution
of superconducting gap structure.\cite{Ota,NXu,Watanabe,XCHong} For the Ba$_{0.1}$K$_{0.9}$Fe$_2$As$_2$ ($T_c$ = 9 K) single crystal,
a special nodal $s$-wave gap structure, with isotropic gap on each $\Gamma$-centered hole pocket
and a ``$\curlyvee$''-shaped nodal gap at the tip of the off-M-centered hole lobes, was revealed by ARPES
experiments.\cite{NXu} This is consistent with the very small residual linear term $\kappa_0/T$ observed by thermal conductivity measurements.\cite{XCHong}
Furthermore, the dramatic doping dependence of $\kappa_0/T$ close to $x$ = 1.0 also can not be explained by a $d$-wave
superconducting gap in KFe$_2$As$_2$.\cite{XCHong}

In this paper, we try to recover the electron Fermi surface in
KFe$_2$As$_2$ alternatively by Co doping, and expect the enhancement
of $T_{\rm c}$ and a crossover from nodal to nodeless
superconducting state. However, unexpected doping effects are found
in K(Fe$_{1-x}$Co$_x$)$_2$As$_2$. First, Co impurities suppress
$T_{\rm c}$ very rapidly, and at only $x$ = 0.042 no
superconductivity can be observed down to 50 mK. This rapid
suppression of $T_{\rm c}$ by impurities is different from that in
Ba$_{0.5}$K$_{0.5}$Fe$_2$As$_2$ with nodeless gaps. Secondly, thermal
conductivity measurements show that the residual linear term $\kappa_0/T$ does
not change much with 3.4\% Co doping. These anomalous impurity effects may reflect
the underlying nodal $s$-wave gap structure in KFe$_2$As$_2$.

K(Fe$_{1-x}$Co$_x$)$_2$As$_2$ single crystals were grown by using
KAs flux method. The starting materials for single crystal growth
are KAs, Fe and Co powders. KAs flux was prepared by reacting
stoichiometric K pieces and As powders at 200 $\celsius$ for 4
hours. KAs, Fe and Co powders were carefully weighed according to
the ratio KAs: Fe: Co=12: 2-2$x$: 2$x$, with $x$ = 0, 0.005, 0.010,
0.025 and 0.050. After thoroughly grounding, the mixtures were put
into alumina crucibles and then sealed in iron crucibles under 1.5
atm of pure argon gas. The sealed crucibles were heated to 900
$\celsius$ in a tube furnace filled with inert gas and kept at 900
$\celsius$ for 6 hours, then cooled slowly to 600 $\celsius$ at 3
$\celsius$/h to grow single crystals. The KAs flux was washed out by
alcohol and shiny black crystals with typical dimensions of 2
$\times$ 1 $\times$ 0.03 mm$^3$ were obtained. The single crystals
are stable in air or alcohol for several days. The actual chemical
compositions were determined by energy dispersive x-ray spectroscopy
to be 0, 0.017, 0.034, 0.042 and 0.050 for the five nominal
compositions, with a standard instrument error 10\%. Below we will
use the actual compositions. Resistivity and Hall coefficient were
measured in a Physical Property Measurement System (PPMS, Quantum
Design). Resistivity below 2 K and thermal conductivity were
measured in a dilution refrigerator.

\begin{figure}[ht]
\centering
\includegraphics[width=0.4\textwidth]{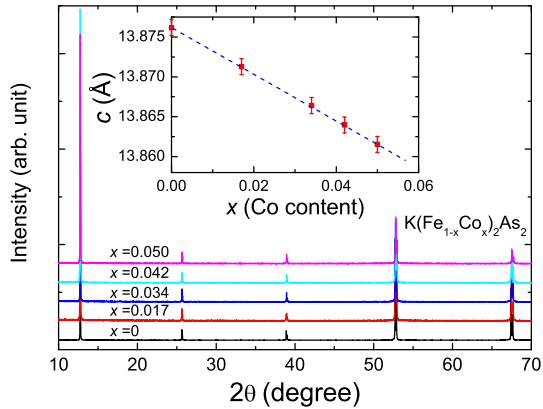}
\caption{(Color online) The XRD patterns for
K(Fe$_{1-x}$Co$_x$)$_2$As$_2$ single crystals with $x$ = 0 - 0.050.
The inset shows $x$ dependence of the lattice parameter $c$
estimated from the data in the main panel.}
\end{figure}

Figure 1 plots the x-ray diffraction (XRD) patterns for
K(Fe$_{1-x}$Co$_x$)$_2$As$_2$ single crystals. Only reflections of
(0 0 2$l$) show up, suggesting good orientation along $c$-axis for
all the crystals. The lattice parameter $c$ was estimated from the
XRD data and plotted in the inset of Fig. 1. For the pure
KFe$_2$As$_2$, $c$ = 13.876 \AA\ is obtained, which is consistent
with previous reports.\cite{Kihou,ChenH} It is found that $c$
decreases linearly with increasing Co doping. At $x$ = 0.050, $c$ is
reduced by $\sim$ 0.1\%.

\begin{figure}[ht]
\centering
\includegraphics[width=0.4\textwidth]{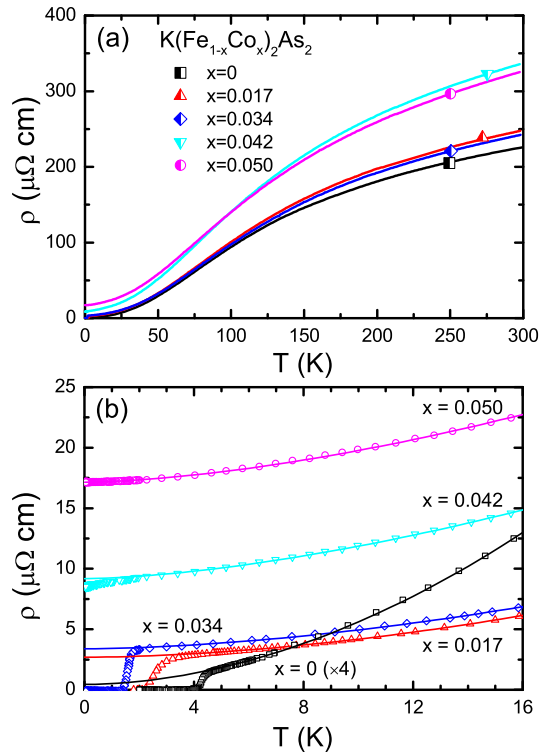}
\caption{(Color online) (a) Temperature dependence of resistivity
for K(Fe$_{1-x}$Co$_x$)$_2$As$_2$ single crystals. (b) The zoomed
plot for the low-temperature data shown in (a). The lines are the
fits to $\rho$ = $\rho_0$ + $aT^\alpha$, from which the residual
resistivity $\rho_0$ is extrapolated.}
\end{figure}

Figure 2 shows the temperature dependence of resistivity for
K(Fe$_{1-x}$Co$_x$)$_2$As$_2$ single crystals. The data were taken
down to 50 mK for $x$ = 0.042 and 0.050 samples. The transition
temperature $T_{\rm c}$ is defined when the resistivity drops to
90\% of the normal-state value. For the pure KFe$_2$As$_2$, $T_{\rm
c}$ = 4.4 K and resistivity reaches zero at 4.2 K. The power-law fit
$\rho(T) = \rho_0 + bT^\alpha$ between 5 and 25 K yields a residual
resistivity $\rho_0$ = 0.11 $\pm$ 0.01 $\mu\Omega$ cm and $\alpha =
1.89$, from which a residual resistance ratio RRR =
$\rho$(300K)/$\rho_0$ = 2017 is estimated. The extremely low
$\rho_0$ and high RRR show that this crystal is very close to the
clean limit. The RRR decreases dramatically from 2017 to 19 with
increasing Co doping from 0 to 0.050 (see Table \ref{1}), suggesting
that the Co doping leads to large impurity scattering. At the same
time, $T_{\rm c}$ decreases rapidly to zero at $x \approx$ 0.04.

\begin{table}
\centering \caption{Lattice parameter $c$, residual resistivity
$\rho_0$, power-law exponent $\alpha$, residual resistance ratio RRR
and superconducting transition temperature $T_{\rm c}$, varying with
Co content $x$.}\label{1}
\begin{tabularx}{0.48\textwidth}{p{1.3cm}p{1.6cm}p{1.6cm}p{1.4cm}p{1.4cm}p{1.1cm}}\hline\hline
$x$ &  $c$ & $\rho_0$   & ~~$\alpha$ & RRR & $T_{\rm c}$ \\
\\ & (\AA)& $(\mu\Omega$ cm) & & & (K) \\ \hline
0    & 13.876 & 0.11   & ~~1.89 &  2017 & 4.4 \\
0.017 & 13.871 & 2.69   & ~~1.79  & 92    & 2.9    \\
0.034 & 13.866 & 3.39   & ~~1.71 &  72 & 1.7     \\
0.042 & 13.864 & 9.18   & ~~1.60 &  37 &  0    \\
0.050 & 13.862 & 17.1 & ~~1.62 &  19 & 0 \\
\hline \hline
\end{tabularx}
\end{table}

Previously, the major role of Co doping in iron-based
superconductors is to introduce electrons into the FeAs
layer, which is evident from the evolution of electron and hole
pockets with $x$ in Ba(Fe$_{1-x}$Co$_x$)$_2$As$_2$.\cite{ChangLiu,MNeupane} The impurity effect of Co dopants is less
concerned. That is why we try to use Co doping to recover the
electron Fermi surfaces in KFe$_2$As$_2$ and expect the increase of
$T_{\rm c}$. However, the rapid suppression of $T_{\rm c}$ with $x$
indicates that the impurity effect of Co dopants is playing the
major role in K(Fe$_{1-x}$Co$_x$)$_2$As$_2$.

Recently, a weak $T_{\rm c}$ suppression effect from
transition-metal impurities was observed in
Ba$_{0.5}$K$_{0.5}$(Fe$_{1-x}$$M_x$)$_2$As$_2$ ($M$ = Mn, Ru, Co,
Ni, Cu, and Zn).\cite{LiJ} In Fig. 3, we plot the normalized
$T_{\rm c}$ versus Co content for K(Fe$_{1-x}$Co$_x$)$_2$As$_2$ and
Ba$_{0.5}$K$_{0.5}$(Fe$_{1-x}$Co$_x$)$_2$As$_2$.\cite{LiJ} For
comparison, similar data of $d$-wave superconductors
YBa$_2$(Cu$_{1-x}$Zn$_x$)$_3$O$_{6.93}$ ($T_{\rm c}$($x$=0) = 92 K),
La$_{1.85}$Sr$_{0.15}$Cu$_{1-x}$Ni$_x$O$_4$ ($T_{\rm
c}$($x$=0) = 37.7 K), and CeCo(In$_{1-x}$Sn$_x$)$_5$ ($T_{\rm
c}$($x$=0) = 2.25 K) are also plotted.\cite{YFukuzumi,CJZhang,MDaniel} It is clearly seen that the
impurity effect on $T_{\rm c}$ in K(Fe$_{1-x}$Co$_x$)$_2$As$_2$ is
very different from that in nodeless $s$-wave superconductor
Ba$_{0.5}$K$_{0.5}$(Fe$_{1-x}$Co$_x$)$_2$As$_2$, but mimics that in
those $d$-wave superconductors.

\begin{figure}[ht]
\centering
\includegraphics[width=0.4\textwidth]{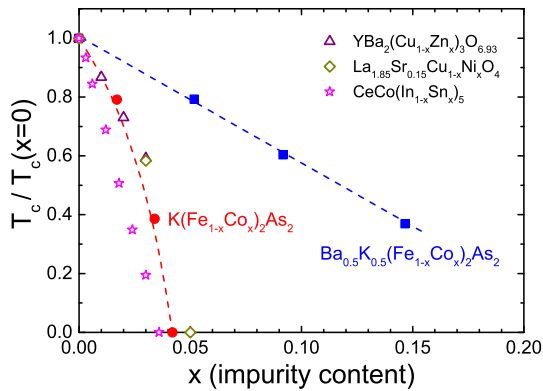}
\caption{(Color online) The normalized $T_{\rm c}$ vs impurity
content $x$ for K(Fe$_{1-x}$Co$_x$)$_2$As$_2$, $s$-wave
superconductor Ba$_{0.5}$K$_{0.5}$(Fe$_{1-x}$Co$_x$)$_2$As$_2$,
$d$-wave superconductors YBa$_2$(Cu$_{1-x}$Zn$_x$)$_3$O$_{6.93}$, La$_{1.85}$Sr$_{0.15}$Cu$_{1-x}$Ni$_x$O$_4$,
and CeCo(In$_{1-x}$Sn$_x$)$_5$.\cite{LiJ,YFukuzumi,CJZhang,MDaniel}}
\end{figure}

The rapid suppression of $T_{\rm c}$ by Co doping in KFe$_2$As$_2$
is very unusual among all iron-based superconductors. As described earlier, recent
ARPES and thermal conductivity experiments support nodal $s$-wave gap in KFe$_2$As$_2$.
Theoretically, for a nodal $s$-wave superconductor, the impurities will make the gap
more isotropic and do not effectively reduce the gap magnitude.\cite{Borkowski} This means 
its $T_{\rm c}$ should not decrease in such a rapid manner. However, recently we also
find similar rapid suppression of $T_{\rm c}$ by Co doping in LiFeP and LaFePO. Therefore,
the anomalous impurity effect on $T_{\rm c}$ may be an intrinsic property of these stoichiometric 
iron-based superconductors with nodal gap. In fact, there was no any real nodal $s$-wave superconductor 
before the discovery of iron-based superconductors. In this sense, further theoretical investigation is needed
to revisit the impurity effect on the $T_{\rm c}$ of nodal $s$-wave superconductor.

An extraordinary feature of $d$-wave superconductor is the universal
heat conduction upon increasing the impurity level, which has been
clearly demonstrated in YBa$_2$(Cu$_{1-x}$Zn$_x$)$_3$O$_{6.9}$.\cite{Taillefer} The universal heat conduction results from the
compensation between the increase of nodal quasiparticle density
induced by impurities and the decrease of mean free path. For an
accidental nodal $s$-wave superconductor, the heat conduction should
not be universal.\cite{VMishra} Reid {\it et al.} observed universal heat conduction by comparing clean and dirty
KFe$_2$As$_2$ single crystals with 10 times difference of $\rho_0$, and argued that it is a compelling evidence for
$d$-wave gap.\cite{Reid} With the intentional Co doping, here we also check whether the heat conduction in K(Fe$_{1-x}$Co$_x$)$_2$As$_2$
is universal.

\begin{figure}[ht]
\centering
\includegraphics[width=0.315\textwidth]{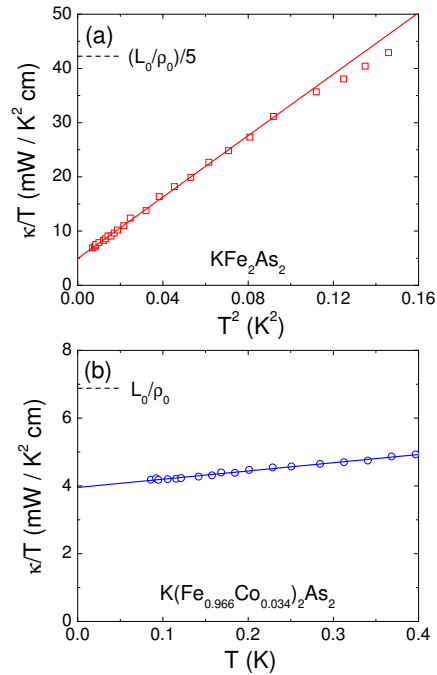}
\caption{(Color online) The thermal conductivity of
K(Fe$_{1-x}$Co$_x$)$_2$As$_2$ with (a) $x$ = 0, and (b) $x$ = 0.034
in zero magnetic field. The solid lines are fits to $\kappa/T = a +
bT^2$ for $x$ = 0 sample, and $\kappa/T = a + bT$ for $x$ = 0.034
sample, respectively. The dash lines are the normal-state
Wiedemann-Franz law expectation $L_0$/$\rho_0$, with $L_0$ the
Lorenz number 2.45 $\times$ 10$^{-8}$ W$\Omega$K$^{-2}$ and
normal-state $\rho_0$ = 0.116 and 3.56 $\mu\Omega$cm, respectively.
For the $x$ = 0 sample, its $L_0$/$\rho_0$ is divided by 5 to put
into the panel.}
\end{figure}

Fig. 4 shows the thermal conductivity of superconducting
K(Fe$_{1-x}$Co$_x$)$_2$As$_2$ single crystals with $x$ = 0 and 0.034
in zero magnetic field. For the pure KFe$_2$As$_2$ with $\rho_0$ =
0.116 $\mu\Omega$ cm, the data below 0.3 K can be fitted by
$\kappa/T = a + bT^2$, with the residual linear term $\kappa_0/T =
a$ = 4.94 $\pm$ 0.09 mW K$^{-2}$ cm$^{-1}$ and $b = 283 \pm 2$.
However, for the $x$ = 0.034 sample with $\rho_0$ = 3.56 $\mu\Omega$
cm, the data below 0.4 K obeys $\kappa/T = a + bT$, with $\kappa_0/T
= a$ = 3.95 $\pm$ 0.01 mW K$^{-2}$ cm$^{-1}$ and $b = 2.43 \pm
0.06$. Previously, for a dirty KFe$_2$As$_2$ with $\rho_0$ = 3.32
$\mu\Omega$ cm, similar temperature dependence of $\kappa/T = a +
bT$ was observed, with $b = 3.04$.\cite{LiSY} Therefore, the $T^2$
term of $\kappa$ is attributed to phonons, and the huge $T^3$ term
observed in the clean KFe$_2$As$_2$ must come from the nodal
quasiparticles. While the origin of this electronic $T^3$ term is still
not clear,\cite{Reid,XCHong} it is apparently suppressed by impurities in the
K(Fe$_{1-x}$Co$_x$)$_2$As$_2$ $x$ = 0.034 sample.

From Fig. 4, while the $\rho_0$ is increased by 30 times, the $\kappa_0/T$ of 
the $x$ = 0 and 0.034 samples in zero field remain comparable (4.94 and 3.95 mW K$^{-2}$ cm$^{-1}$). 
For the $x$ = 0 sample, the value is less than 3\% of its normal-state
Wiedemann-Franz law expectation $\kappa/T$ = $L_0$/$\rho_0$, with
$L_0$ the Lorenz number 2.45 $\times$ 10$^{-8}$ W$\Omega$K$^{-2}$.
For the dirty $x$ = 0.034 sample, the value is more that 50\% of its
normal-state $\kappa/T$. Therefore, roughly universal heat conduction is also
observed in K(Fe$_{1-x}$Co$_x$)$_2$As$_2$. Although the theoretical model with some assumed nodal $s$-wave gap structure
suggests that the thermal conductivity is nonuniversal,\cite{VMishra} it is hard to conclude that this comparable heat conduction 
is an evidence of $d$-wave gap in KFe$_2$As$_2$.\cite{Reid} First, the superconducting gap structure of KFe$_2$As$_2$ is very complex, 
e.g., the ``$\curlyvee$''-shaped nodal gap at the tip of the off-M-centered hole lobes. Second, the impurity may slightly
change the doping level, which can also cause a dramatic change of $\kappa_0/T$.\cite{XCHong} One needs to carefully evaluate whether 
different effects can finally bring some kind of compensation on thermal conductivity in KFe$_2$As$_2$.

In summary, we have grown high-quality K(Fe$_{1-x}$Co$_x$)$_2$As$_2$
single crystals and find that the superconductivity was suppressed
quickly by Co doping at only $x \approx$ 0.04. Such an effective
suppression of $T_{\rm c}$ by impurities is very different from that
in nodeless $s$-wave superconductor Ba$_{0.5}$K$_{0.5}$Fe$_2$As$_2$, 
and it may be an intrinsic property of stoichiometric
iron-based superconductors with nodal gap. A comparable heat conduction
is observed in the $x$ = 0 and 0.034 superconducting samples, however, 
a careful evaluation of different effects on $\kappa_0/T$  is needed, taking into account
the complex nodal $s$-wave superconducting gap structure of KFe$_2$As$_2$.

This work is supported by National Natural Science Foundation of
China (Grant No. 11190021, 51021091, and 91021016), the National
Basic Research Program of China (973 Program, Grant No.
2012CB922002, 2011CB00101, and 2012CB821402), Chinese Academy of
Sciences, and the Program for Professor of Special Appointment
(Eastern Scholar) at Shanghai Institutions of Higher Learning.

\end{document}